\documentclass[aps,prl,twocolumn,letterpaper,superscriptaddress,showpacs]{revtex4}
\usepackage{keyval}%
\usepackage{graphicx}
\usepackage{dcolumn}
\usepackage{bm}
\usepackage{color}
\usepackage{times,xspace}
\usepackage{amsbsy,amssymb,amsmath}
\usepackage{graphicx,color,epsfig,rotate}
\usepackage{fancyhdr}

\setkeys{Gin}{width=0.8\columnwidth,clip=true}

\newcommand{\ignore}[1]{}

\newcommand{\ket}[1]{|{#1}\rangle}

\newcommand{\tp}{t^\prime}
\newcommand{\tpp}{t^{\prime\prime}}

\newcommand{\epz}{\varepsilon^{}_{Pz}}
\newcommand{\tcmax}{T^\mathrm{max}_c}
\newcommand{\veff}{V_{ij}}
\newcommand{\ueff}{\mu_{i}}
\newcommand{\ezero}{\varepsilon_{ij}}

\ignore{
\pagestyle{fancy} %
\pagestyle{fancyplain} %
\lhead{\large Yin \& Ku} %
\chead{\sl submitted to PHYSICAL REVIEW LETTERS\vspace{-2pt}}
\rhead{\today} %
\cfoot{\sc\thepage} %
\lfoot{} %
\rfoot{} %
}

\begin{document}


\title{Tuning Hole Mobility, Concentration, and Repulsion in High-$T_c$ Cuprates via Apical Atoms}

\author{Wei-Guo Yin}
\affiliation{Condensed Matter Physics \& Materials Science Department, Brookhaven National Laboratory, Upton, NY 11973} %
\author{Wei Ku}
\affiliation{Condensed Matter Physics \& Materials Science Department, Brookhaven National Laboratory, Upton, NY 11973} %
\affiliation{Physics Department, State University of New York, Stony Brook, NY 11790} %

\date{Received \today }

\begin{abstract}
Using a newly developed first-principles Wannier-states approach
that takes into account large on-site Coulomb repulsion, we derive
the effective low-energy interacting Hamiltonians for several
prototypical high-$T_c$ superconducting cuprates. The material
dependence is found to originate primarily from the different energy
of the apical atom $p_z$ state. Specifically, the general properties
of the low-energy hole state, namely the Zhang-Rice singlet, are
significantly modified by a triplet state associated with this $p_z$
state, via additional intra-sublattice hoppings, nearest-neighbor
``super-repulsion'', and other microscopic many-body processes.
Possible implications on modulation of $T_c$, local superconducting
gaps, charge distribution, hole mobility, electron-phonon
interaction, and multi-layer effects are discussed.
\end{abstract}
\maketitle

\textsl{Introduction.}---The origin of high-$T_c$ superconductivity
(HTSC) remains under fierce debate notwithstanding monumental effort
for two decades \cite{HTC:review:cho}.  This fascinating phenomenon
is achieved when a number of layered copper oxides are doped away
from their Mott insulating parent phase. Although it is generally
agreed that the most relevant electron behavior is confined within
the common (CuO$_2$)$^{2-}$ plane, $T_c^\mathrm{max}$ ($T_c$ at
optimal doping) strikingly varies from 28 K in
Ca$_{2-x}$Na$_x$CuO$_2$Cl$_2$ to 135 K in
HgBa$_2$Ca$_2$Cu$_3$O$_{8+\delta}$ by modulation of the layering
pattern along the less essential third direction. Hence, clarifying
the material dependence of the in-plane electron behavior,
especially in unbiased first-principles-based approaches, is an
essential and effective step toward the resolution of the HTSC
mechanism and the quest for higher $T_c$ superconductors.

A recent influential advancement along the line was made by Andersen
and coworkers \cite{HTC:apical:pavarini}. Within the local-density
approximation (LDA) of density-functional theory, they derived a
one-band \emph{noninteracting} Hamiltonian for several hole-doped
cuprates, showing that $-\tp/t$ ($t$ and $\tp$ are the first- and
second-nearest-neighbor hopping integrals, respectively) was
strongly material-dependent and correlated with $T_c^\mathrm{max}$.
Taken as the kinetic part of the effective one-band $t$-$J$ or
Hubbard model, the most studied model for the CuO$_2$ plane
\cite{HTC:review:anderson_04}, this single band has been widely used
to compare with angular-resolved photoemission spectroscopy (ARPES)
\cite{HTC:review:arpes,HTC:la2cuo4:arpes:tanaka,HTC:yin_prl98}.

A particular puzzle thus caused is that the LDA results of
$-\tp/t=0.18$ for La$_2$CuO$_4$ and $0.12$ for Ca$_2$CuO$_2$Cl$_2$
\cite{HTC:apical:pavarini} are in striking contradiction to their
diamond- and square-like Fermi surfaces, respectively, as observed
in ARPES \cite{HTC:review:arpes,HTC:la2cuo4:arpes:tanaka}. In
addition, the low-energy carriers in hole-doped cuprates are well
accepted to be closely related to the Zhang-Rice singlet (ZRS) that
naturally appears in deriving the $t$-$J$ model from an interacting
many-body picture \cite{HTC:model:ZRS}. This two-body singlet state
is \emph{intrinsically different} from any single-particle state.
Therefore, it is essential to reexamine the material dependence by
moving one step forward to derive an effective \emph{interacting}
Hamiltonian in a first-principles approach that takes into account,
from the beginning, the large Coulomb repulsion on the Cu sites. New
material-dependent interaction effects are then expected to be
revealed from the strong electronic interaction and the
charge-transfer nature \cite{HTC:note:CT}.

In this Letter, using a recently developed first-principles
Wannier-states (WSs) approach \cite{lamno3:yin_prl06,wannier:ku}
which takes into account large on-site Coulomb repulsion, we derive
the effective low-energy interacting Hamiltonians for several
prototypical high-$T_c$ superconducting cuprates. The only
significant microscopic difference among the cuprates is found to
originate from the apical atom $p_z$ state. Strong impact of the
apical atom on $\tp$ is observed; in particular, $-\tp/t$ for
Ca$_2$CuO$_2$Cl$_2$ is found to be considerably larger than for
La$_2$CuO$_4$, in good agreement with ARPES. Moreover, we show that
besides the $\tp$ and electrostatic effects, the realistic variation
of the energy level of the apical $p_z$ states is able to tune
substantially the local site potential and intersite
``super-repulsion,'' directly modulating the local superconducting
gap and charge distribution. This provides a natural explanation of
recent spectroscopic imaging scanning tunneling microscopy (SI-STM)
observations \cite{HTC:bi2212:STM:mcelroy,HTC:bi2212:STM:davis}.
Finally, implication on a new realization of the electron-phonon
coupling is briefly addressed. These important findings should shed
new light on the general material dependence and microscopic
understanding of HTSC.

\textsl{Methods.}---A three-step approach is employed to
systematically reduce the energy scale of the relevant Hilbert
space: (i) The full-energy electronic structures are obtained within
the LDA$+U$ approximation \cite{HTC:note:ldau} known as a
state-of-the-art generalization of LDA to include strong local
interaction. (ii) At the intermediate-energy scale ($\sim 10$ eV
covering the Cu $3d$, O $2p$, and relevant apical orbitals), an
effective five-band interacting Hamiltonian, $H^{5b}$, is derived
\cite{lamno3:yin_prl06}. The local part of $H^{5b}$, which includes
the leading terms (on-Cu-site interaction and local $d$-$p$
hybridization), is then diagonalized for all the doping levels
\cite{\ignore{theory:hubbard_65,}HTC:apical:raimondi,HTC:model:mattis_95,HTC:model:kostyrko}.
(iii) At the low-energy scale ($\sim 1$ eV), an effective one-band
Hamiltonian, $H^{1b}$, is derived from canonical transformation to
project out high-energy states up to the second order
\cite{HTC:model:kostyrko}.

Since the apical (out-of-plane) coordination of the planar Cu
cations is the main structural variation relevant to the physics of
the CuO$_2$ plane, we present here a study of a number of cuprates
representing different apical limits: La$_2$CuO$_4$ with two apical
oxygen atoms per copper site, Ca$_2$CuO$_2$Cl$_2$ with the apical
$p_z$ energy level, $\epz$, moved away from the Fermi level by
substitution of Cl for O, Sr$_2$CuO$_2$F$_2$ with $\epz$ moved
farther away, and Nd$_2$CuO$_4$ without apical atoms (equivalently,
$\epz=\infty$).

\begin{figure}[b]
\includegraphics[width=0.95\columnwidth,clip=true]{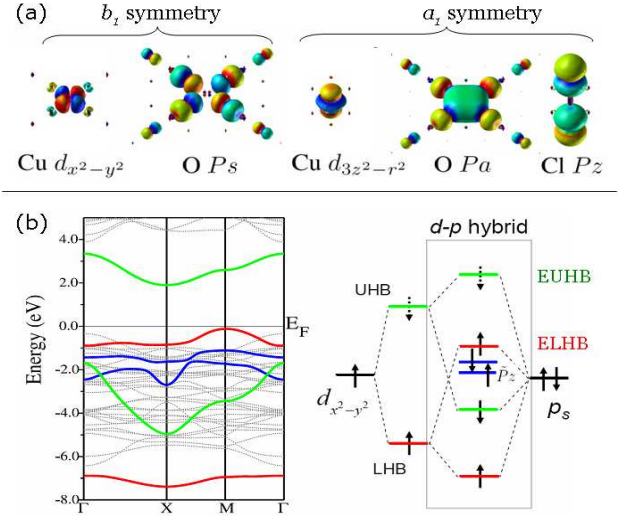}
\caption{\label{fig:BS}%
(Color online) (a) Wannier functions centered at Cu sites for
$H^{5b}$: O $P_s$ and $P_a$ are constructed by maximizing the weight
of the $b_1$- and $a_1$-symmetrized combination of four neighboring
planar O $p_\sigma$ orbitals, respectively; $P_z$ is the
$a_1$-symmetrized combination of two apical $p_z$ orbitals. (b) Left
panel: Band structures of Ca$_2$CuO$_2$Cl$_2$ from LDA$+U$
calculations (dots) and Wannier states analysis (lines).
$\Gamma$=$(0,0,0)$, $X$=$(\pi,0,0)$,
$M$=$(\frac{\pi}{2},\frac{\pi}{2},0)$. Right panel: Corresponding
local energy level splitting due to interactions and $d$-$p$
hybridization, filled with electrons (solid arrows) or holes
(dashed arrows). $\tilde{P}z$ are derived from two highest $a_1$ hybrids.}%
\end{figure}

The most relevant WSs are shown in Fig.~\ref{fig:BS}(a), including
the Cu $3d$, O $2p$, and apical $p$ orbitals, spanning an energy
window of $\sim 10$ eV [c.f. Fig.~\ref{fig:BS}(b)]. Since these WSs
almost completely decouple from other WSs, a five-band Hamiltonian
$H^\mathrm{5b}$ with on-site Coulomb and exchange interactions among
the Cu $d_{x^2-y^2}$ and $d_{3z^2-r^2}$ orbitals can be
unambiguously derived by matching its self-consistent Hartree-Fock
expression with the Wannier representation of the LDA$+U$
Hamiltonian \cite{lamno3:yin_prl06}. The details will be presented
elsewhere. Indeed, \emph{among the cuprates, the most noticeable
difference in the parameters of $H^\mathrm{5b}$ is $\epz$}
\cite{HTC:note:epz}, as presented in Table~\ref{table:t-J}.

The likelihood of a one-band picture is readily revealed by our WSs
analysis of the LDA$+U$ band. Fig.~\ref{fig:BS}(b) shows such an
analysis for Ca$_2$CuO$_2$Cl$_2$, for there exist excellent ARPES
data to compare with. Note that the lowest unoccupied band is mainly
of Cu $3d_{x^2-y^2}$ character and the highest occupied band is
mainly of O $P_s$ character. Apparently, these two in-plane bands
are the most relevant to the low-energy physics; indeed, their
overall shapes agree well with the ARPES \cite{HTC:review:arpes}. In
addition, their corresponding WSs have opposite spin characters, as
clearly shown in the right panel of Fig.~\ref{fig:BS}(b). Thus, the
simplest one-band picture is to view these two anti-spin-polarized
bands (both are antibonding between Cu $3d_{x^2-y^2}$ and O $P_s$
WSs) as the \emph{effective} upper and lower Hubbard bands (EUHB and
ELHB) with the charge-transfer gap as the effective Hubbard gap.

A more appropriate approach is to project out high-energy states
systematically, based on exact diagonalization of the leading terms
(here the \emph{local} part of $H^\mathrm{5b}$)
\cite{\ignore{theory:hubbard_65,}HTC:apical:raimondi,HTC:model:mattis_95,HTC:model:kostyrko}.
Specifically, our approach benefits greatly from having all the
\emph{first-principles} WSs, including the O $2p$ states, centered
at the Cu sites [see Fig.~\ref{fig:BS}(a)] by construction. This
reduces the decorated CuO$_2$ lattice to a simple square lattice,
and unambiguously defines the local Hamiltonian to be diagonalized.

It is appropriate to keep in the low-energy space only the local
one-hole and two-hole ground states, referred to as $\ket{\sigma}$
and $\ket{\mathrm{ZRS}}$, respectively, as far as the mechanism of
HTSC is concerned, for which the relevant energy scale is expected
to be smaller than $0.5$ eV \cite{HTC:review:anderson_04}. Note that
the first-excited two-hole state is a spin-triplet (referred to as
$\ket{\mathrm{axial}}$) composed of one $b_1$ hole and one $a_1$
hole with excitation energy $\Delta_2=0.81$ eV for
Ca$_2$CuO$_2$Cl$_2$, for example. A one-band model is thus obtained
by projecting out all the other states by canonical perturbation and
then mapping to a constrained fermion system without double
occupancy \cite{HTC:model:kostyrko}.

Additional caution was paid to La$_2$CuO$_4$. With a model
tetragonal structure
\cite{HTC:apical:pavarini,HTC:la2cuo4:LDA:mcmahan_88,HTC:la2cuo4:LDA:hybertsen_90},
we obtain $\epz-\varepsilon^{}_{Ps}=-0.52$ eV and $\Delta_2$ is
merely $0.13$ eV, suggesting that a two-band Hamiltonian including
$\ket{\mathrm{axial}}$ be more appropriate. For this subtle case,
other more realistic considerations are important; for example, we
find that the $5^\circ$ octahedral-tilting in the real material
increases $\epz-\varepsilon^{}_{Ps}$ by $0.25$ eV and $\Delta_2$ by
100\%. Such a tilting effect is interesting, as the oxygen dopants
in Bi$_2$Sr$_2$CaCu$_2$O$_{8+\delta}$ also induce the tilting of
nearby oxygen octahedra. To account for such effects, we present in
Table~\ref{table:t-J} the results for the model La$_2$CuO$_4$ with
$\epz-\varepsilon^{}_{Ps}$ being simply increased by 0.8 eV (now
$\Delta_2=0.5$ eV); the $\epz$ effects are still large though
considerably abated.

\vspace{0.2cm} %
\textsl{Results and Discussion.}---The resulting effective one-band
$t$-$J$-like Hamiltonian is given by \cite{HTC:note:t-J}
\begin{eqnarray}
 H^\mathrm{1b} &=& -\sum\limits_{ {ij} \sigma }
 {t_{ij} (\tilde c_{i\sigma }^\dag  \tilde c_{j\sigma }^{}  + H.c.)}
 + J\sum\limits_{\left\langle {ij} \right\rangle } {(\vec{S}_i^{}  \cdot \vec{S}_j^{}  - \frac{{n_i n_j }}{4})}
 \nonumber \\
  &+& \frac{J}{4}\sum\limits_{\left\langle {ijk} \right\rangle \sigma }
{(\tilde c_{i\sigma }   ^\dag  \tilde c_{j\bar \sigma }^\dag \tilde
c_{j\sigma }^{}  \tilde c_{k\bar \sigma }^{}  - \tilde c_{i\sigma
}^\dag  n_{j\bar \sigma }^{} \tilde c_{k\sigma }^{}  + H.c.)}
  \nonumber  \\
&+& \sum\limits_{\left\langle {ij} \right\rangle } \veff{n_i n_j } +
\sum\limits_{i} \ueff{n_i} + \sum_{ij}{\ezero}, \label{eq:t-J}
\end{eqnarray}
where $\tilde c_{i\sigma }^{}$ is the annihilation operator for the
constrained fermion with spin $\sigma$ at site $i$. $n_{i\sigma}
=\tilde c_{i\sigma }^\dag \tilde c_{i\sigma }^{}$,
$n_{i}=\sum_\sigma{n_{i\sigma}}$, and $\vec{S}_i = \sum_{\mu\nu}
\tilde c_{i\sigma }^\dag \vec{\sigma}_{\mu\nu} \tilde
c_{i\bar{\sigma} }^{}$ with $\vec{\sigma}_{\mu\nu}$ being the Pauli
matrices. All the site pairs $\left\langle {ij} \right\rangle$ stand
for nearest neighbors except $t_{ij}$ extends to the third nearest
neighbors (referred to as $\tpp$). Note that the first two lines of
Eq.~(\ref{eq:t-J}) resemble the $t$-$J$ model mapped out from the
one-band Hubbard model and have been extensively studied
\cite{HTC:review:dagotto_94}. The other terms include
$\epz$-dependent ``super-repulsion'' $\veff$, site potential
$\ueff$, and energy ``constant'' $\ezero$.

The derived parameters for several prototypical cuprates upon hole
doping are listed in Table~\ref{table:t-J}. For comparison, the
contributions purely from the $b_1$ orbitals (from which ZRS is
formed) are given inside the parentheses in Table~\ref{table:t-J}.
Remarkably, \emph{there exists practically no material dependence if
only the $b_1$ orbitals are considered}. The universality for $t$
and $J$ survives inclusion of the $a_1$ orbitals, since they are
solely determined by the $b_1$ orbitals, suggesting that the
magnetic mechanism be responsible for the robustness of
superconductivity among the cuprates.

\begin{table}[b]
\caption{\label{table:t-J}%
Derived parameters of Eq.~(\ref{eq:t-J}) for four prototypical
cuprates upon hole doping. Inside the parentheses are contributions
purely from the $b_1$ orbitals. The energy unit is meV.}
\begin{ruledtabular}
\begin{tabular}{lccccc}
 & Nd$_2$CuO$_4$ & Sr$_2$CuO$_2$F$_2$ & Ca$_2$CuO$_2$Cl$_2$  & La$_2$CuO$_4$\\
\hline %
$T_c^\mathrm{max}$ (K)     & N/A      & 46 & 28 & 38 \\
$\epz-\varepsilon^{}_{Ps}$ & $\infty$ & 3010 & 810 & 280\footnotemark[1] \\
$J_{\mathrm{LDA}+U}$\footnotemark[2]    & 131          & 155 &  145           &   144           \\
$J/2$             & 138 (138) &  138 (138) & 131 (131) & 129 (129)\\
$t$        &  431 (431)        & 467 (467)   &  459 (459)  &   488 (488) \\
$\tp/|t|$  & -0.33 (-0.35) &-0.25 (-0.32) & -0.19 (-0.32)   &   0.01 (-0.30) \\
$\tpp/|t|$ &  0.23 (0.24) & 0.19 (0.23) &  0.16 (0.23) &   0.06 (0.22) \\
$\veff$              & 29 (27) & 39 (29) &  54 (30)  &  137 (31)      \\
$\ueff$              & -795 (-800) & -337 (-394) &  -295 (-412)  &  -354 (-529)      \\
$\ezero$              & -36 (-14) & -86 (-17) &  -133 (-17)  &  -258 (-19)      \\
\end{tabular}
\end{ruledtabular}
\footnotetext[1]{Increased by $0.8$ eV (see text).}%
\footnotetext[2]{Estimated from total energies of the (anti)ferromagnetic states.}%
\end{table}

\begin{figure}[b]
\includegraphics[width=0.95\columnwidth,clip=true]{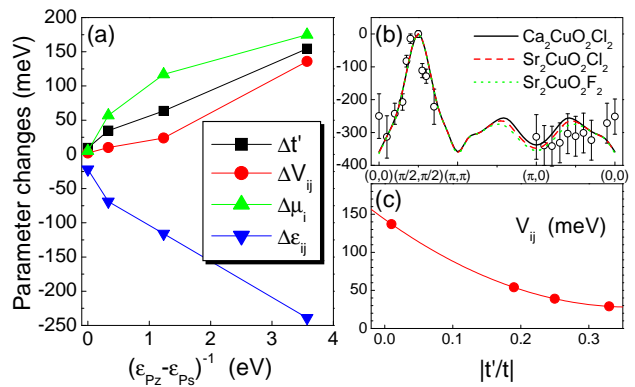}
\caption{\label{fig:pz}%
(a) Parameter changes by the $a_1$ orbitals as a function of $\epz$.
(b) Calculated one-hole quasiparticle dispersions compared with
ARPES on Sr$_2$CuO$_2$Cl$_2$ (open circles) \cite{HTC:review:arpes}.
(c) $\veff$ versus $|\tp/t|$.}%
\end{figure}

\begin{figure}[b]
\includegraphics[width=0.95\columnwidth,clip=true]{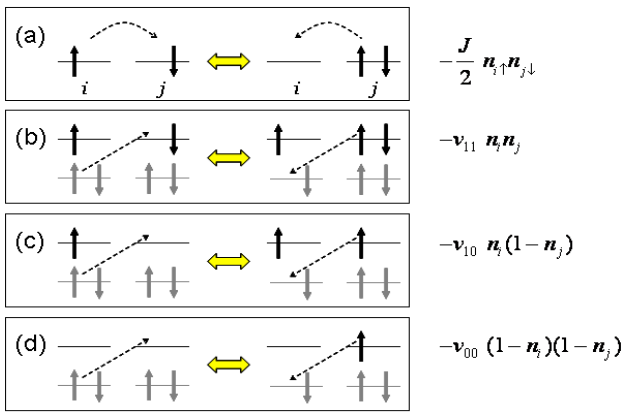}
\caption{\label{fig:vaccuum}%
Schematics of virtual kinematical processes and corresponding
effective one-band interaction terms for (a) Heisenberg
superexchange within a one-orbital system and (b)-(d) additional
``vacuum fluctuations'' in a two-orbital system.
Electrons (solid arrows) in the targeted and projected out orbitals
are black and gray, respectively. Dashed lines denote hopping paths.}%
\end{figure}

More amazingly, the material dependence of these materials
\emph{almost entirely results from the influence of the apical $P_z$
orbitals}; the other $a_1$ orbitals have negligible contributions as
made clear from comparing all the cases with Nd$_2$CuO$_4$. $\tp$,
$\tpp$, $\veff$, $\ueff$, and $\ezero$ all show strong material-
($\epz$-) dependence [c.f. Fig.~\ref{fig:pz}(a)]. First of all,
three-site kinematical processes via virtual nearest-neighbor
hoppings to the $P_z$ orbital dramatically renormalize $\tp$ and
$\tpp$, as suggested in Ref. \cite{HTC:apical:raimondi}. These
modulations would seriously affect the mobility of hole
quasiparticles and the shape of the Fermi surface, given that the
nearest-neighbor hoppings of hole quasiparticles are suppressed by
the antiferromagnetic spin correlation \cite{HTC:review:dagotto_94}.
Note that $-\tp/t$ is substantially smaller in La$_2$CuO$_4$ than in
Ca$_2$CuO$_2$Cl$_2$, consistent with their observed diamond- and
square-like Fermi surfaces, respectively
\cite{HTC:review:arpes,HTC:la2cuo4:arpes:tanaka}.

Quantitatively, our calculated values of $-\tp/t$ for the above two
compounds are considerably smaller than the widely used ones ($\sim
0.15$ and $0.3$, respectively) estimated from fitting ARPES in the
very lightly doped $t$-$J$ model
\cite{HTC:review:arpes,HTC:la2cuo4:arpes:tanaka,HTC:yin_prl98,HTC:sr2cuo2cl2:arpes:xiang}.
Interestingly, the single-hole quasiparticle dispersions calculated
with the present parameters for $A$$_2$CuO$_2$X$_2$ ($A$=Ca, Sr;
$X$=F, Cl) in the self-consistent Born approximation (SCBA)
\cite{HTC:yin_prl98} still agree well with available ARPES data
\cite{HTC:review:arpes}, as shown in Fig.~\ref{fig:pz}(b). This is
because the dispersion is actually more sensitive to $\tpp/t$, which
is consistent between ours and previous estimations. Moreover, the
fitted values of $\tp$ and $\tpp$ were shown to depend sensitively
on inclusion of the three-site hopping terms $\propto J/4$
\cite{HTC:sr2cuo2cl2:arpes:xiang}. Our first-principles derived
Hamiltonian finally provides an unambiguous benchmark of these
material-dependent parameters.

In sharp contrast, the ``downfolded'' LDA one-band led to a puzzling
opposite trend ($-\tp/t=0.18$ for La$_2$CuO$_4$ and $0.12$ for
Ca$_2$CuO$_2$Cl$_2$) \cite{HTC:apical:pavarini} that appears in
contradiction with ARPES and our parameters. This is, however, not
alarming, since $t$'s in Eq. (1) describes hoppings of ZRS
(intersite swapping of the $\ket{\mathrm{ZRS}}$ and $\ket{\sigma}$
states \cite{HTC:model:ZRS}), while the LDA $t$'s describes hopping
of the single-particle LDA $\ket{\sigma}$ state. This intrinsic
difference can be easily observed from Fig.~\ref{fig:BS}(b) where
EUHB and ELHB are shown to have \emph{different} WSs characters, an
essential feature missed in the picture of the LDA one-band plus
effective Hubbard interaction \cite{HTC:apical:pavarini}. Clearly,
building the strong many-body characteristics in first-principles
approaches before the downfolding
\cite{HTC:la2cuo4:LDA:mcmahan_88,HTC:la2cuo4:LDA:hybertsen_90}, as
presented here, is necessary for the low-energy physics of cuprates.

Next, the striking appearance of the unconventional ``super-repulsion''
$\veff$ originates from two-site kinematical processes as demonstrated
in Fig.~\ref{fig:vaccuum}.
For a purely one-band system in the large on-site repulsion limit
[Fig.~\ref{fig:vaccuum}(a)], the virtual hoppings between two
singly-occupied neighboring sites generate the well-known
superexchange effect $-\frac{J}{2} n_{i\sigma}n_{j\bar{\sigma}}$.
Given an extra fully-occupied orbital to be projected out
($P_z$ in the present work), additional virtual kinematical processes
give rise to three new spin-independent terms [$\propto v_{00}$,
$v_{10}$, and $v_{11}$, see Fig.~\ref{fig:vaccuum}(b)-(d)] to the
targeted one-band Hamiltonian, leading to $\veff=2v_{10}-v_{00}-v_{11}$,
$\ueff=4(v_{00}-v_{10})$ and $\ezero=-v_{00}$.

This new effective repulsion, $\veff$ (named ``super-repulsion'' in
analog to superexchange), is \emph{to be distinguished from direct
Coulomb interaction}: Not only is $\veff$ controlled by $\epz$, but
also its virtual kinematical origin makes it less subject to the
electronic screening. Besides, as presented in Fig.~\ref{fig:pz}(c),
$\veff$ rapidly increases as $|\tp/t|$ decreases from Nd$_2$CuO$_2$
to La$_2$CuO$_2$. Therefore, for any realistic study of the
cuprates, the $\veff$ effect should be addressed along with the
$\tp$ effect.

As for superconductivity, the repulsive $\veff$ apparently weakens
the local pairing strength, regardless of the actual pairing
mechanism, given that paired holes tend to reside as nearest
neighbors (implied by the $d_{x^2-y^2}$-wave symmetry of the order
parameter \cite{HTC:review:tsuei}). This provides a natural avenue
for modulating the local superconducting gap
\cite{HTC:bi2212:STM:mcelroy} as well as $T_c$. On the other hand,
the suggested trend of $\tcmax \sim \tp/t$
~\cite{HTC:apical:raimondi,HTC:apical:pavarini}---which turned out
to be controversial in several numerical studies
\cite{HTC:t':white_99,HTC:t':maier_00,HTC:t':shih_04}---is clearly
violated by comparing Ca$_{2-x}$Na$_x$CuO$_2$Cl$_2$
($\tcmax$=$28$~K) with La$_{2-x}$Sr$_x$CuO$_4$ ($\tcmax$=$38$~K). It
is thus fair to speculate that the full material dependence of
$\tcmax$ involves other complex aspects of real materials (e.g.,
sample cleanness, doping approaches, two-band scenarios). Further
investigation is needed to identify those aspects that affect
$\tcmax$ beyond our present discovery.

With regard to charge distribution (CD), modulation of $\ueff$ tends
to induce charge inhomogeneity. As shown in Fig.~\ref{fig:pz}(a),
$\ueff$ depends strongly on the apical environment.
This explains strong \emph{interlayer} charge inhomogeneity in
multilayer systems such as
HgBa$_2$Ca$_{n-1}$Cu$_n$O$_{2n+2+\delta}$ with $n \geq 3$
\cite{HTC:Hg1245:mukuda} and atomically perfect
La$_{1.85}$Sr$_{0.15}$CuO$_4$/La$_2$CuO$_4$ films
\cite{HTC:la2cuo4:bozovic}, as $\ueff$ in distinct CuO$_2$ layers
can be considerably different. On the other hand, repulsive $\veff$,
which disfavors hole accumulation,
effectively counters $\ueff$ 
to give a smoother \emph{in-plane} CD.

The present results would shed new light on the understanding of
other exotic features of cuprates. For example, recent SI-STM
experiments on Bi$_2$Sr$_2$CaCu$_2$O$_{8+\delta}$
indicated that local superconducting gaps vary from $20-70$~meV
correlating with oxygen dopants, O$_\delta$, \emph{while} low-energy
charge density variations are weak \cite{HTC:bi2212:STM:mcelroy}.
These striking results have been phenomenologically attributed to a
strong local modulation of electron pairing strength
\cite{HTC:bi2212:model:nunner}, $J$ in the $t$-$J$ model
\cite{HTC:bi2212:model:zhu}, or $\ueff$ with two types of O$_\delta$
\cite{HTC:bi2212:model:zhou}. The present studies suggest that
O$_\delta$ perturb O$_\mathrm{apical}$ and thus significantly
modulate $\veff$, instead of $J$, in addition to $\ueff$, $\tp$, and
$\tpp$. This new scenario also agrees well with more recent SI-STM
measurements that reveal a strong correlation between the local
superconducting gap and Cu--O$_\mathrm{apical}$ distance
\cite{HTC:bi2212:STM:davis}.

Furthermore, the strong $\epz$ dependence of $\tp$, $\tpp$, $\veff$,
and $\ueff$ implies the electron-phonon interactions involving
vibration of the apical atoms may be strong. In particular, the
$\veff$ variation in this way yields an unconventional higher-order
electron-phonon interaction. The determination of their actual
strengths will be pursued elsewhere.
The enhanced electron-phonon interactions could cooperate with
$\veff$ in favoring other competing orders such as ``stripe,''
especially in doped La$_2$CuO$_4$
\cite{HTC:la2cuo4:stripe:tranquada} where we have shown that the
pair-breaking $V_{ij}$
is comparable to $J$. 

Finally, our findings point to the importance of examining HTSC in
multi-layer systems \cite{HTC:Hg1245:mukuda,HTC:F0234:chen} or
cuprate superlattices \cite{HTC:la2cuo4:bozovic}, as the influence
of the apical $P_z$ orbitals will be significantly layer-dependent
and can be controlled (e.g., via strain and field effects) to tune
hole concentration, mobility, and super-repulsion among the CuO$_2$
layers, a very interesting problem for future investigation.

We are grateful to S.R. White for his advice on numerical canonical
transformation \cite{Algorithm:white_02}. We thank J.C. Davis for
presenting us SI-STM results \cite{HTC:bi2212:STM:davis} prior to
publication. Helpful
discussions with O.K. Andersen, P.W. Anderson, I. Bozovic, 
T.K. Lee, D.C. Mattis, W. E. Pickett, T.M. Rice, G. Sawatzky, R.
Scalettar, A.M. Tsvelik, and Z. Wang are also acknowledged.
This work was supported by US DOE (DE-AC02-98CH10886) and DOE-CMSN.


\begin{thebibliography}{34}
\expandafter\ifx\csname
natexlab\endcsname\relax\def\natexlab#1{#1}\fi
\expandafter\ifx\csname bibnamefont\endcsname\relax
  \def\bibnamefont#1{#1}\fi
\expandafter\ifx\csname bibfnamefont\endcsname\relax
  \def\bibfnamefont#1{#1}\fi
\expandafter\ifx\csname citenamefont\endcsname\relax
  \def\citenamefont#1{#1}\fi
\expandafter\ifx\csname url\endcsname\relax
  \def\url#1{\texttt{#1}}\fi
\expandafter\ifx\csname urlprefix\endcsname\relax\def\urlprefix{URL
}\fi \providecommand{\bibinfo}[2]{#2}
\providecommand{\eprint}[2][]{\url{#2}}

\bibitem[{\citenamefont{Cho}(2006)}]{HTC:review:cho}
\bibinfo{author}{\bibfnamefont{A.}~\bibnamefont{Cho}},
  \bibinfo{journal}{Science} \textbf{\bibinfo{volume}{314}},
  \bibinfo{pages}{1072} (\bibinfo{year}{2006}).

\bibitem[{\citenamefont{Pavarini et~al.}(2001)\citenamefont{Pavarini, Dasgupta,
  Saha-Dasgupta, Jepsen, and Andersen}}]{HTC:apical:pavarini}
\bibinfo{author}{\bibfnamefont{E.}~\bibnamefont{Pavarini}},
  \bibinfo{author}{\bibfnamefont{I.}~\bibnamefont{Dasgupta}},
  \bibinfo{author}{\bibfnamefont{T.}~\bibnamefont{Saha-Dasgupta}},
  \bibinfo{author}{\bibfnamefont{O.}~\bibnamefont{Jepsen}}, \bibnamefont{and}
  \bibinfo{author}{\bibfnamefont{O.~K.} \bibnamefont{Andersen}},
  \bibinfo{journal}{Phys.\ Rev.\ Lett.} \textbf{\bibinfo{volume}{87}},
  \bibinfo{pages}{047003} (\bibinfo{year}{2001}).

\bibitem[{\citenamefont{Anderson et~al.}(2004)\citenamefont{Anderson, Lee,
  Randeria, Rice, Trivedi, and Zhang}}]{HTC:review:anderson_04}
\bibinfo{author}{\bibfnamefont{P.~W.} \bibnamefont{Anderson}},
  \bibinfo{author}{\bibfnamefont{P.~A.} \bibnamefont{Lee}},
  \bibinfo{author}{\bibfnamefont{M.}~\bibnamefont{Randeria}},
  \bibinfo{author}{\bibfnamefont{T.~M.} \bibnamefont{Rice}},
  \bibinfo{author}{\bibfnamefont{N.}~\bibnamefont{Trivedi}}, \bibnamefont{and}
  \bibinfo{author}{\bibfnamefont{F.~C.} \bibnamefont{Zhang}},
  \bibinfo{journal}{J. Phys.: Condens. Matter} \textbf{\bibinfo{volume}{16}},
  \bibinfo{pages}{R755} (\bibinfo{year}{2004}).

\bibitem[{\citenamefont{Damascelli et~al.}(2003)\citenamefont{Damascelli, Shen,
  and Hussain}}]{HTC:review:arpes}
\bibinfo{author}{\bibfnamefont{A.}~\bibnamefont{Damascelli}},
  \bibinfo{author}{\bibfnamefont{Z.-X.} \bibnamefont{Shen}}, \bibnamefont{and}
  \bibinfo{author}{\bibfnamefont{Z.}~\bibnamefont{Hussain}},
  \bibinfo{journal}{Rev.\ Mod.\ Phys.} \textbf{\bibinfo{volume}{75}},
  \bibinfo{pages}{473} (\bibinfo{year}{2003}).

\bibitem[{\citenamefont{Tanaka et~al.}(2004)\citenamefont{Tanaka, Yoshida,
  Fujimori, Lu, Shen, Zhou, Eisaki, Hussain, Uchida, Aiura
  et~al.}}]{HTC:la2cuo4:arpes:tanaka}
\bibinfo{author}{\bibfnamefont{K.}~\bibnamefont{Tanaka}},
  \bibinfo{author}{\bibfnamefont{T.}~\bibnamefont{Yoshida}},
  \bibinfo{author}{\bibfnamefont{A.}~\bibnamefont{Fujimori}},
  \bibinfo{author}{\bibfnamefont{D.~H.} \bibnamefont{Lu}},
  \bibinfo{author}{\bibfnamefont{Z.-X.} \bibnamefont{Shen}},
  \bibinfo{author}{\bibfnamefont{X.-J.} \bibnamefont{Zhou}},
  \bibinfo{author}{\bibfnamefont{H.}~\bibnamefont{Eisaki}},
  \bibinfo{author}{\bibfnamefont{Z.}~\bibnamefont{Hussain}},
  \bibinfo{author}{\bibfnamefont{S.}~\bibnamefont{Uchida}},
  \bibinfo{author}{\bibfnamefont{Y.}~\bibnamefont{Aiura}},
  \bibnamefont{et~al.}, \bibinfo{journal}{Phys.\ Rev.\ B}
  \textbf{\bibinfo{volume}{70}}, \bibinfo{pages}{092503}
  (\bibinfo{year}{2004}).

\bibitem[{\citenamefont{Yin et~al.}(1998)\citenamefont{Yin, Gong, and
  Leung}}]{HTC:yin_prl98}
\bibinfo{author}{\bibfnamefont{W.-G.} \bibnamefont{Yin}},
  \bibinfo{author}{\bibfnamefont{C.-D.} \bibnamefont{Gong}}, \bibnamefont{and}
  \bibinfo{author}{\bibfnamefont{P.~W.} \bibnamefont{Leung}},
  \bibinfo{journal}{Phys.\ Rev.\ Lett.} \textbf{\bibinfo{volume}{81}},
  \bibinfo{pages}{2534} (\bibinfo{year}{1998}).

\bibitem[{\citenamefont{Zhang and Rice}(1988)}]{HTC:model:ZRS}
\bibinfo{author}{\bibfnamefont{F.~C.} \bibnamefont{Zhang}} \bibnamefont{and}
  \bibinfo{author}{\bibfnamefont{T.~M.} \bibnamefont{Rice}},
  \bibinfo{journal}{Phys. Rev. B} \textbf{\bibinfo{volume}{37}},
  \bibinfo{pages}{3759} (\bibinfo{year}{1988}).

\bibitem[{HTC({\natexlab{a}})}]{HTC:note:CT}
\bibinfo{note}{Related discussions can be found in M. V. Mostovoy and D. I.
  Khomskii, Phys. Rev. Lett. \textbf{92}, 167201 (2004).}

\bibitem[{\citenamefont{Yin et~al.}(2006)\citenamefont{Yin, Volja, and
  Ku}}]{lamno3:yin_prl06}
\bibinfo{author}{\bibfnamefont{W.-G.} \bibnamefont{Yin}},
  \bibinfo{author}{\bibfnamefont{D.}~\bibnamefont{Volja}}, \bibnamefont{and}
  \bibinfo{author}{\bibfnamefont{W.}~\bibnamefont{Ku}},
  \bibinfo{journal}{Phys.\ Rev.\ Lett.} \textbf{\bibinfo{volume}{96}},
  \bibinfo{pages}{116405} (\bibinfo{year}{2006}).

\bibitem[{\citenamefont{Ku et~al.}(2002)\citenamefont{Ku, Rosner, Pickett, and
  Scalettar}}]{wannier:ku}
\bibinfo{author}{\bibfnamefont{W.}~\bibnamefont{Ku}},
  \bibinfo{author}{\bibfnamefont{H.}~\bibnamefont{Rosner}},
  \bibinfo{author}{\bibfnamefont{W.~E.} \bibnamefont{Pickett}},
  \bibnamefont{and} \bibinfo{author}{\bibfnamefont{R.~T.}
  \bibnamefont{Scalettar}}, \bibinfo{journal}{Phys. Rev. Lett.}
  \textbf{\bibinfo{volume}{89}}, \bibinfo{pages}{167204}
  (\bibinfo{year}{2002}).

\bibitem[{\citenamefont{McElroy et~al.}(2005)\citenamefont{McElroy, Lee,
  Slezak, Lee, Eisaki, Uchida, and Davis}}]{HTC:bi2212:STM:mcelroy}
\bibinfo{author}{\bibfnamefont{K.}~\bibnamefont{McElroy}},
  \bibinfo{author}{\bibfnamefont{J.}~\bibnamefont{Lee}},
  \bibinfo{author}{\bibfnamefont{J.~A.} \bibnamefont{Slezak}},
  \bibinfo{author}{\bibfnamefont{D.-H.} \bibnamefont{Lee}},
  \bibinfo{author}{\bibfnamefont{H.}~\bibnamefont{Eisaki}},
  \bibinfo{author}{\bibfnamefont{S.}~\bibnamefont{Uchida}}, \bibnamefont{and}
  \bibinfo{author}{\bibfnamefont{J.~C.} \bibnamefont{Davis}},
  \bibinfo{journal}{Science} \textbf{\bibinfo{volume}{309}},
  \bibinfo{pages}{1048} (\bibinfo{year}{2005}).

\bibitem[{\citenamefont{Davis}()}]{HTC:bi2212:STM:davis}
\bibinfo{author}{\bibfnamefont{J.~C.} \bibnamefont{Davis}},
  \bibinfo{note}{unpublished}.

\bibitem[{HTC({\natexlab{b}})}]{HTC:note:ldau}
\bibinfo{note}{We applied the WIEN2k [P. Blaha \textsl{et al.}, Comput. Phys.
  Commun. \textbf{147}, 71 (2002)] implementation of the full potential
  linearized augmented plane wave method in the LDA$+U$ approach with $U=8$ eV
  and $J=0.88$ eV [V.I. Anisimov \textsl{et al.}, Phys. Rev. B \textbf{70},
  172501 (2004) and references therein].}

\bibitem[{\citenamefont{Raimondi et~al.}(1996)\citenamefont{Raimondi,
  Jefferson, and Feiner}}]{HTC:apical:raimondi}
\bibinfo{author}{\bibfnamefont{R.}~\bibnamefont{Raimondi}},
  \bibinfo{author}{\bibfnamefont{J.~H.} \bibnamefont{Jefferson}},
  \bibnamefont{and} \bibinfo{author}{\bibfnamefont{L.~F.}
  \bibnamefont{Feiner}}, \bibinfo{journal}{Phys.\ Rev.\ B}
  \textbf{\bibinfo{volume}{53}}, \bibinfo{pages}{8774} (\bibinfo{year}{1996}).

\bibitem[{\citenamefont{Mattis}(1995)}]{HTC:model:mattis_95}
\bibinfo{author}{\bibfnamefont{D.~C.} \bibnamefont{Mattis}},
  \bibinfo{journal}{Phys.\ Rev.\ Lett.} \textbf{\bibinfo{volume}{74}},
  \bibinfo{pages}{3676} (\bibinfo{year}{1995}).

\bibitem[{\citenamefont{Kostyrko}(1989)}]{HTC:model:kostyrko}
\bibinfo{author}{\bibfnamefont{T.}~\bibnamefont{Kostyrko}},
  \bibinfo{journal}{Phys.\ Rev.\ B} \textbf{\bibinfo{volume}{40}},
  \bibinfo{pages}{4596} (\bibinfo{year}{1989}).

\bibitem[{HTC({\natexlab{c}})}]{HTC:note:epz}
\bibinfo{note}{The strong material dependence of $\epz$ was also found in the
  classic ionic model [Y. Ohta, T. Tohyama, and S. Maekawa, Phys.\ Rev.\ B
  \textbf{43}, 2968 (1991)].}

\bibitem[{\citenamefont{McMahan et~al.}(1988)\citenamefont{McMahan, Martin, and
  Satpathy}}]{HTC:la2cuo4:LDA:mcmahan_88}
\bibinfo{author}{\bibfnamefont{A.~K.} \bibnamefont{McMahan}},
  \bibinfo{author}{\bibfnamefont{R.~M.} \bibnamefont{Martin}},
  \bibnamefont{and} \bibinfo{author}{\bibfnamefont{S.}~\bibnamefont{Satpathy}},
  \bibinfo{journal}{Phys.\ Rev.\ B} \textbf{\bibinfo{volume}{38}},
  \bibinfo{pages}{6650} (\bibinfo{year}{1988}).

\bibitem[{\citenamefont{Hybertsen et~al.}(1990)\citenamefont{Hybertsen,
  Stechel, Schluter, and Jennison}}]{HTC:la2cuo4:LDA:hybertsen_90}
\bibinfo{author}{\bibfnamefont{M.~S.} \bibnamefont{Hybertsen}},
  \bibinfo{author}{\bibfnamefont{E.~B.} \bibnamefont{Stechel}},
  \bibinfo{author}{\bibfnamefont{M.}~\bibnamefont{Schluter}}, \bibnamefont{and}
  \bibinfo{author}{\bibfnamefont{D.~R.} \bibnamefont{Jennison}},
  \bibinfo{journal}{Phys.\ Rev.\ B} \textbf{\bibinfo{volume}{41}},
  \bibinfo{pages}{11068} (\bibinfo{year}{1990}).

\bibitem[{HTC({\natexlab{d}})}]{HTC:note:t-J}
\bibinfo{note}{Small corrections of three-site hopping terms including
  $c^\dag_i(1-n_j)c^{}_k$ have been neglected.}

\bibitem[{\citenamefont{Dagotto}(1994)}]{HTC:review:dagotto_94}
\bibinfo{author}{\bibfnamefont{E.}~\bibnamefont{Dagotto}},
  \bibinfo{journal}{Rev.\ Mod.\ Phys.} \textbf{\bibinfo{volume}{66}},
  \bibinfo{pages}{763} (\bibinfo{year}{1994}).

\bibitem[{\citenamefont{Xiang and Wheatley}(1996)}]{HTC:sr2cuo2cl2:arpes:xiang}
\bibinfo{author}{\bibfnamefont{T.}~\bibnamefont{Xiang}} \bibnamefont{and}
  \bibinfo{author}{\bibfnamefont{J.~M.} \bibnamefont{Wheatley}},
  \bibinfo{journal}{Phys.\ Rev.\ B} \textbf{\bibinfo{volume}{54}},
  \bibinfo{pages}{R12653} (\bibinfo{year}{1996}).

\bibitem[{\citenamefont{Tsuei and Kirtley}(2000)}]{HTC:review:tsuei}
\bibinfo{author}{\bibfnamefont{C.~C.} \bibnamefont{Tsuei}} \bibnamefont{and}
  \bibinfo{author}{\bibfnamefont{J.~R.} \bibnamefont{Kirtley}},
  \bibinfo{journal}{Rev.\ Mod.\ Phys.} \textbf{\bibinfo{volume}{72}},
  \bibinfo{pages}{969} (\bibinfo{year}{2000}).

\bibitem[{\citenamefont{White and Scalapino}(1999)}]{HTC:t':white_99}
\bibinfo{author}{\bibfnamefont{S.~R.} \bibnamefont{White}} \bibnamefont{and}
  \bibinfo{author}{\bibfnamefont{D.~J.} \bibnamefont{Scalapino}},
  \bibinfo{journal}{Phys.\ Rev.\ B} \textbf{\bibinfo{volume}{60}},
  \bibinfo{pages}{R753} (\bibinfo{year}{1999}).

\bibitem[{\citenamefont{Maier et~al.}(2000)\citenamefont{Maier, Jarrell,
  Pruschke, and Keller}}]{HTC:t':maier_00}
\bibinfo{author}{\bibfnamefont{T.}~\bibnamefont{Maier}},
  \bibinfo{author}{\bibfnamefont{M.}~\bibnamefont{Jarrell}},
  \bibinfo{author}{\bibfnamefont{T.}~\bibnamefont{Pruschke}}, \bibnamefont{and}
  \bibinfo{author}{\bibfnamefont{J.}~\bibnamefont{Keller}},
  \bibinfo{journal}{Phys.\ Rev.\ Lett.} \textbf{\bibinfo{volume}{85}},
  \bibinfo{pages}{1524} (\bibinfo{year}{2000}).

\bibitem[{\citenamefont{Shih et~al.}(2004)\citenamefont{Shih, Lee, Eder, Mou,
  and Chen}}]{HTC:t':shih_04}
\bibinfo{author}{\bibfnamefont{C.~T.} \bibnamefont{Shih}},
  \bibinfo{author}{\bibfnamefont{T.~K.} \bibnamefont{Lee}},
  \bibinfo{author}{\bibfnamefont{R.}~\bibnamefont{Eder}},
  \bibinfo{author}{\bibfnamefont{C.-Y.} \bibnamefont{Mou}}, \bibnamefont{and}
  \bibinfo{author}{\bibfnamefont{Y.~C.} \bibnamefont{Chen}},
  \bibinfo{journal}{Phys.\ Rev.\ Lett.} \textbf{\bibinfo{volume}{92}},
  \bibinfo{pages}{227002} (\bibinfo{year}{2004}).

\bibitem[{\citenamefont{Mukuda et~al.}(2006)\citenamefont{Mukuda, Abe, Araki,
  Kitaoka, Tokiwa, Watanabe, Iyo, Kito, and Tanaka}}]{HTC:Hg1245:mukuda}
\bibinfo{author}{\bibfnamefont{H.}~\bibnamefont{Mukuda}},
  \bibinfo{author}{\bibfnamefont{M.}~\bibnamefont{Abe}},
  \bibinfo{author}{\bibfnamefont{Y.}~\bibnamefont{Araki}},
  \bibinfo{author}{\bibfnamefont{Y.}~\bibnamefont{Kitaoka}},
  \bibinfo{author}{\bibfnamefont{K.}~\bibnamefont{Tokiwa}},
  \bibinfo{author}{\bibfnamefont{T.}~\bibnamefont{Watanabe}},
  \bibinfo{author}{\bibfnamefont{A.}~\bibnamefont{Iyo}},
  \bibinfo{author}{\bibfnamefont{H.}~\bibnamefont{Kito}}, \bibnamefont{and}
  \bibinfo{author}{\bibfnamefont{Y.}~\bibnamefont{Tanaka}},
  \bibinfo{journal}{Phys.\ Rev.\ Lett.} \textbf{\bibinfo{volume}{96}},
  \bibinfo{pages}{087001} (\bibinfo{year}{2006}).

\bibitem[{\citenamefont{Bozovic et~al.}(2003)\citenamefont{Bozovic, Logvenov,
  Verhoeven, Caputo, Goldobin, and Geballe}}]{HTC:la2cuo4:bozovic}
\bibinfo{author}{\bibfnamefont{I.}~\bibnamefont{Bozovic}},
  \bibinfo{author}{\bibfnamefont{G.}~\bibnamefont{Logvenov}},
  \bibinfo{author}{\bibfnamefont{M.~A.~J.} \bibnamefont{Verhoeven}},
  \bibinfo{author}{\bibfnamefont{P.}~\bibnamefont{Caputo}},
  \bibinfo{author}{\bibfnamefont{E.}~\bibnamefont{Goldobin}}, \bibnamefont{and}
  \bibinfo{author}{\bibfnamefont{T.~H.} \bibnamefont{Geballe}},
  \bibinfo{journal}{Nature} \textbf{\bibinfo{volume}{422}},
  \bibinfo{pages}{873} (\bibinfo{year}{2003}).

\bibitem[{\citenamefont{Nunner et~al.}(2005)\citenamefont{Nunner, Andersen,
  Melikyan, and Hirschfeld}}]{HTC:bi2212:model:nunner}
\bibinfo{author}{\bibfnamefont{T.~S.} \bibnamefont{Nunner}},
  \bibinfo{author}{\bibfnamefont{B.~M.} \bibnamefont{Andersen}},
  \bibinfo{author}{\bibfnamefont{A.}~\bibnamefont{Melikyan}}, \bibnamefont{and}
  \bibinfo{author}{\bibfnamefont{P.~J.} \bibnamefont{Hirschfeld}},
  \bibinfo{journal}{Phys.\ Rev.\ Lett.} \textbf{\bibinfo{volume}{95}},
  \bibinfo{pages}{177003} (\bibinfo{year}{2005}).

\bibitem[{\citenamefont{Zhu}()}]{HTC:bi2212:model:zhu}
\bibinfo{author}{\bibfnamefont{J.-X.} \bibnamefont{Zhu}},
  \bibinfo{note}{cond-mat/0508646}.

\bibitem[{\citenamefont{Zhou et~al.}(2007)\citenamefont{Zhou, Ding, and
  Wang}}]{HTC:bi2212:model:zhou}
\bibinfo{author}{\bibfnamefont{S.}~\bibnamefont{Zhou}},
  \bibinfo{author}{\bibfnamefont{H.}~\bibnamefont{Ding}}, \bibnamefont{and}
  \bibinfo{author}{\bibfnamefont{Z.}~\bibnamefont{Wang}},
  \bibinfo{journal}{Phys.\ Rev.\ Lett.} \textbf{\bibinfo{volume}{98}},
  \bibinfo{pages}{076401} (\bibinfo{year}{2007}).

\bibitem[{\citenamefont{Tranquada et~al.}(2004)\citenamefont{Tranquada, Woo,
  Perring, Goka, Gu, Xu, Fujita, and Yamada}}]{HTC:la2cuo4:stripe:tranquada}
\bibinfo{author}{\bibfnamefont{J.~M.} \bibnamefont{Tranquada}},
  \bibinfo{author}{\bibfnamefont{H.}~\bibnamefont{Woo}},
  \bibinfo{author}{\bibfnamefont{T.~G.} \bibnamefont{Perring}},
  \bibinfo{author}{\bibfnamefont{H.}~\bibnamefont{Goka}},
  \bibinfo{author}{\bibfnamefont{G.~D.} \bibnamefont{Gu}},
  \bibinfo{author}{\bibfnamefont{G.}~\bibnamefont{Xu}},
  \bibinfo{author}{\bibfnamefont{M.}~\bibnamefont{Fujita}}, \bibnamefont{and}
  \bibinfo{author}{\bibfnamefont{K.}~\bibnamefont{Yamada}},
  \bibinfo{journal}{Nature} \textbf{\bibinfo{volume}{429}},
  \bibinfo{pages}{534} (\bibinfo{year}{2004}).

\bibitem[{\citenamefont{Chen et~al.}(2006)\citenamefont{Chen, Iyo, Yang, Zhou,
  Lu, Eisaki, Devereaux, Hussain, and Shen}}]{HTC:F0234:chen}
\bibinfo{author}{\bibfnamefont{Y.}~\bibnamefont{Chen}},
  \bibinfo{author}{\bibfnamefont{A.}~\bibnamefont{Iyo}},
  \bibinfo{author}{\bibfnamefont{W.}~\bibnamefont{Yang}},
  \bibinfo{author}{\bibfnamefont{X.}~\bibnamefont{Zhou}},
  \bibinfo{author}{\bibfnamefont{D.}~\bibnamefont{Lu}},
  \bibinfo{author}{\bibfnamefont{H.}~\bibnamefont{Eisaki}},
  \bibinfo{author}{\bibfnamefont{T.~P.} \bibnamefont{Devereaux}},
  \bibinfo{author}{\bibfnamefont{Z.}~\bibnamefont{Hussain}}, \bibnamefont{and}
  \bibinfo{author}{\bibfnamefont{Z.-X.} \bibnamefont{Shen}},
  \bibinfo{journal}{Phys.\ Rev.\ Lett.} \textbf{\bibinfo{volume}{97}},
  \bibinfo{pages}{236401} (\bibinfo{year}{2006}).

\bibitem[{\citenamefont{White}(2002)}]{Algorithm:white_02}
\bibinfo{author}{\bibfnamefont{S.~R.} \bibnamefont{White}},
  \bibinfo{journal}{J.\ Chem.\ Phys.} \textbf{\bibinfo{volume}{117}},
  \bibinfo{pages}{7472} (\bibinfo{year}{2002}).

\end{thebibliography}

\end{document}